\documentclass[useAMS,usenatbib]{mn2e}

\usepackage{natbib, aas_macros}
\usepackage[dvips]{graphicx}
\usepackage{wrapfig}
\usepackage{color}
\usepackage{amsmath}
\usepackage{amssymb}

\newcommand{\Msun}{M_{\odot}}

\newcommand{\Msunyr}{{\rm M_{\odot}\;yr^{-1}}}
\newcommand{\Lsun}{L_{\odot}}

\title[Sub-millimeter brightness of early star-forming galaxies]
{Sub-millimeter brightness of early star-forming galaxies}
\author[Yajima et al.]
{Hidenobu Yajima$^{1}$\thanks{E-mail: yuh19@psu.edu(HY); 
umemura@ccs.tsukuba.ac.jp(MU); mmori@ccs.tsukuba.ac.jp(MM)}, 
Masayuki Umemura$^{2}$, 
Masao Mori$^{2}$
\\
$^{1}$Department of Astronomy and Astrophysics, Pennsylvania State University,
525 Davey Lab, University Park, PA 16802, USA\\
$^{2}$Center for Computational Sciences, University of Tsukuba, 
Tsukuba 305-8577, Japan
}
 
\begin{document}

\date{Accepted ?; Received ??; in original form ???}

\pagerange{\pageref{firstpage}--\pageref{lastpage}} \pubyear{2011}

\maketitle

\label{firstpage}

%
%
\begin{abstract}

Based on a three-dimensional model of an early star-forming galaxy,
we explore the evolution of the sub-millimeter brightness.
The model galaxy is employed from an ultra-high-resolution chemodynamic 
simulation of a primordial galaxy by Mori \& Umemura,
where the star formation rate (SFR) is $\sim 10 \rm\; \Msunyr$ 
at $t_{\rm age} \lesssim 0.3$ Gyr and several $\rm\; \Msunyr$ at 
$t_{\rm age} > 0.3$ Gyr. The former phase well reproduces the observed 
properties of Lyman alpha emitters (LAEs) and the latter does
Lyman break galaxies (LBGs). 
We solve the three-dimensional radiative transfer in the clumpy
interstellar media in this model galaxy, taking the size distributions 
of dust grains into account, and calculate the dust temperature as a function
of galactic evolutionary time. 
We find that the clumpiness of interstellar media plays an important role
for the sub-millimeter brightness. 
In the LAE phase, dust grains are concentrated on clumpy 
star-forming regions that are distributed all over the galaxy, 
and the grains can effectively absorb UV radiation from stars.
As a result, the dust is heated up to $T_{\rm dust} \gtrsim 35$ K.
In the LBG phase, the continuous supernovae drive dust grains far away 
from star-forming regions.
Then, the grains cannot absorb much radiation from stars,
and becomes into a cold state close to the CMB temperature.
Consequently, the dust temperature decreases with the evolutionary time,
where the mass-weighted mean temperature is 
$T_{\rm dust} = 26~\rm K$ at $t_{\rm age}=$ 0.1 Gyr and 
$T_{\rm dust} = 21~\rm K$ at $t_{\rm age}=$ 1.0 Gyr.
By this analysis, it turns out that the sub-millimeter brightness is higher
in the LAE phase than that in the LBG phase, although the dust-to-gas ratio
increases monotonically as a function of time. 
We derive the spectral energy distributions by placing the model
galaxy at a given redshift.  
The peak flux at 850 $\rm \mu m$ is found to be $S_{850} \sim 0.2 - 0.9$ mJy 
if the model galaxy is placed at $6 \geq z \geq 2$.
This means that ALMA can detect an early star-forming galaxy 
with SFR of $\sim 10 \rm\; \Msunyr$ by less than one hour integration 
with 16 antennas. 
\end{abstract}

%
%
\begin{keywords}
radiative transfer -- ISM: dust, extinction -- galaxies: evolution -- galaxies: formation -- galaxies: high-redshift -- methods: numerical
\end{keywords}

%
%
\section{Introduction}

Exploring the properties of early star-forming galaxies is one of key themes
to elucidate the galaxy formation process. 
The fashion of star formation in galaxies can significantly change
dust properties. In young galaxies, type-II supernovae create dust,
and also change the dust size and component \citep{Dwek98, Todini01, Bianchi07, Nozawa06-1, Nozawa07}.
In addition, UV radiation from young stars is effectively absorbed by dust
and therefore determines the temperature of dust grains. 
Besides, the dust can affect the star formation efficiency itself
via hydrogen molecule formation and the cooling by thermal emission.
Accordingly, dust properties such as amount, temperature, size and composition 
should tightly correlate with the star formation history.
Hence, understanding the dust properties is a significant issue
in the study of galaxy evolution.

So far, the sub-millimeter (sub-mm) flux from distant galaxies 
has been detected at redshifts of $0.3-4$. 
These galaxies are called sub-mm galaxies (SMGs)
\citep[e.g.,][]{Smail97-1, Hughes98, Barger98, Eales99, Ivison00, Daddi09}.
However, the observed galaxies are limited to extremely luminous starburst galaxies,
where the star formation rate (SFR) is $\sim 1000~ \rm \Msunyr$,
from the restriction by the detection limit of the facility
\citep[e.g.,][]{Borys03, Greve04, Chapman05, Laurent05, Coppin06,  Bertoldi07,  Perera08, Scott08, Weis09,  Austermann10, Eales10, Hatsukade11}. 
Hence, the dust properties in normal star-forming galaxies 
(SFR $\lesssim 100~ \rm \Msunyr$) at high redshifts 
(which mean $z \gtrsim 3$ in this paper)
have not been hitherto understood well.

Recently, some observations are dedicated to detect the sub-mm flux 
from early star-forming galaxies
(SFR $\sim 0.1 - 100~ \rm \Msunyr$) like
Lyman alpha emitters (LAEs) and Lyman break galaxies (LBGs) by using newly developed facilities
\citep{Tamura09, Stanway10}.
Although \citet{Tamura09} did not detect notable sub-mm flux from LAEs and LBGs 
in the SSA22 region within the detection limit of $\rm S_{1100} \sim 2~ mJy$,
they discovered sub-mm sources in the vicinity of LAEs.
\citet{Stanway10} showed that the upper-limit of sub-mm flux for LBGs at $z \sim 5$ 
is $\rm S_{870}< 0.85\; mJy$ from stacked data. 
Hence, with recent sub-mm telescopes (e.g. Hershel, AzTEK), 
it seems very difficult to detect sub-mm flux from high-z star-forming galaxies 
with SFR$< 100~ \rm \Msunyr$. 
Moreover, the determination of  dust temperature is a hard task
in the observational study,
because it requires the flux data at many wavelengths in sub-mm band.
Hence, one often assumes the dust temperature to speculate
spectral energy distributions (SED).
Then, the estimation of $L_{\rm IR}$ by integrating speculated SED 
suffer from the uncertainty of the assumed dust temperature.
The uncertainty leads to the misestimation of SFR and dust amount.
A theoretical model may help to relate the observed sub-mm flux with
the dust properties in star-forming galaxies. 
Such attempts have been made by several authors 
\citep[e.g.,][]{Takagi03-1, Dayal10-1}.
\citet{Takagi03} constructed an SED model of dusty
star-forming galaxies by solving the radiative transfer in spherical symmetry. 
\citet{Dayal10}, based on cosmological SPH simulations, 
employed homogeneous distributions of stars and dust in an identified LAE.
Recently, \citet{Mori06} have shown through ultra-high-resolution numerical 
simulations that interstellar media in early star-forming galaxies become quite 
clumpy as a result of multiple supernovae.
Hence, to construct the SED for early star-forming galaxies, 
it is imperative to solve 
three-dimensional radiative transfer in clumpy interstellar media.

Here, we theoretically explore the dust amount and temperature 
in star-forming galaxies by solving three-dimensional radiative transfer,
based on a chemodynamic simulation of a primordial galaxy 
by \citet{Mori06}. 
Since the model galaxy well reproduces the observed properties of 
Lyman alpha emitters (LAEs) at $t_{\rm age} \lesssim 0.3$ Gyr 
and Lyman break galaxies (LBGs) 
at $t_{\rm age} > 0.3$ Gyr, we can predict the sub-mm brightness in the
LAE and LBG phases in early star-forming galaxies.
In the present analyses,
the cosmological parameters are assumed to be 
$H_{\rm 0}=70$ km s$^{-1}$ Mpc$^{-1}$, 
$\Omega_{\rm M}=0.3$ and $\Omega_{\rm \Lambda}=0.7$.
In \S 2, the model and numerical method are described.
In \S 3, the results on dust properties in early star-forming galaxies
are presented.
In \S 4, the detectability by ALMA is discussed.
\S 5 is devoted to the summary.

%
%
\section{Model \& Method}

\subsection{Model galaxy}
Our model galaxy is a supernova-dominated star-forming galaxy 
simulated by an ultra-high-resolution
($1024^{3}$ fixed Cartesian grids) chemodynamics calculation coupled
with the collisionless dynamics of star particles. 
The simulation pursues the early evolution ($< 2\times 10^{9}$ years)
of a primeval galaxy as an assemblage of sub-galactic condensations of 
$5.0 \times 10^{9} ~\Msun$, 
building up a system with the total mass of $10^{11} ~\Msun$
[see \citet{Mori06} for the simulation details].

The star formation rate (SFR) of the model galaxy is 
$11 \rm\; \Msunyr$ at $t_{\rm age}= 0.1$ Gyr, 
$10 \rm\; \Msunyr$ at $t_{\rm age}= 0.3$ Gyr,
$8 \rm\; \Msunyr$ at $t_{\rm age}= 0.5$ Gyr, and 
$5 \rm\; \Msunyr$ at $t_{\rm age}= 1$ Gyr.
The metal and dust in the galaxy is enriched by the continuous type II 
supernovae due to vigorous star formation, 
and exhibits complex inhomogeneous dust distributions \citep[figure 1 in][]{Yajima09}.

According to the Lyman $\alpha$ luminosity, 
the early evolutionary stage is divided into two phases:
one is the LAE-phase ($t_{\rm age} \lesssim 0.3$ Gyr) 
and the other is the LBG-phase ($t_{\rm age} > 0.3$ Gyr).
Most of Lyman $\alpha$ emission comes from the cooling radiation 
by interstellar gas, and the luminosity 
reaches $2.0\times10^{43}$ erg s$^{-1}$ at $t_{\rm age} = 0.1$ Gyr 
and $1.6\times10^{43}$ erg s$^{-1}$ at $t_{\rm age}=0.3$ Gyr,
respectively. They nicely match the observed luminosity of LAEs
and also well resemble LAEs with respect to other properties. 
At $t_{\rm age}\geq 0.5$ Gyr, the Lyman $\alpha$ luminosity quickly declines 
to several $10^{41}$ erg s$^{-1}$ that is lower than the detection limit,
and the SED is dominated by stellar continuum. This phase appears like LBGs.
The model galaxy at this phase has stellar age ($\sim 2-6 \times 10^{8}~\rm yr$) 
and mass ($\sim 6-9 \times 10^{9}~\Msun$) similar to typical LBGs at $z \sim 3$,
while it is older and more massive than typical LBGs at $z \gtrsim 5$ \citep{Verma07}.
On the other hand, a part of LBGs at $z \gtrsim 5$ have been detected by {\it Spitzer} IRAC,
and show the similar age and mass to our model galaxy \citep{Eyles07, Stark09}.
Therefore, our model galaxy at a later phase is probably corresponding 
to typical LBGs for $z \sim 3$,
and a massive subset of LBGs for $z \gtrsim 5$.

\subsection{Radiative transfer}
\label{sec:radiation}

The radiation from young stars propagates in the 
highly inhomogeneous interstellar media containing dust.
We compute the three-dimensional radiative transfer (RT)
of stellar radiation to derive the dust temperature.
The RT calculations are done as the post-processing for each
evolutionary stage of model galaxy. For the purpose, the data 
of the hydrodynamic simulations are coarse-grained into $128^3$ Cartesian grids.

The RT scheme used in this paper is the Authentic Radiation Transfer (ART) 
method which is originally developed by \citet{Nakamoto01-1}. 
The procedure is basically the same as that in \citet{Yajima09, Yajima11}. 
Our ART method is based on the {\it long-characteristic} method.
Usually, the {\it short-characteristic} method is computationally less costly
than the long-characteristic method, by an order of $N$ which is a grid number 
in the linear dimension.
However, the short-characteristic method suffers from numerical diffusion effect.
The ART method is devised to reduce the computational cost to a level similar 
to the short-characteristic method with keeping the accuracy equivalent to 
the long-characteristic method.
Hence, the present method allows us to calculate the transfer of radiation from 
a large number of sources.

In this work, 
the RT equation is solved along $16384$ rays with uniform angular resolution 
from each star particle. 
The number of star particles is $3 - 8.5 \times 10^{4}$. 
Hence, we carry out the RT calculation for $\sim 10^{8-9}$ rays for each snapshot.
We map individual star particles to nearest grids,
and set radiation rays in an isotropic fashion from each star particle.

\subsection{Dust model}
\label{sec:dust}
We distribute the interstellar dust in proportion to the metallicity,
assuming the size distribution of $dn_{\rm d} / da_{\rm d} \propto a^{-3.5}_{\rm d}$
\citep{Mathis77-1}, where $a_{\rm d}$ is the radius of a dust grain.
We suppose the grain size in the range of $0.1 -1.0\,\mu$m as our fiducial model.
The dust mass in a cell is calculated by a following simple relation 
between metallicity and dust \citep{Draine07},
\begin{equation}
 m_{\rm d}=0.01m_{\rm g} \frac{Z}{Z_{\odot}}, 
\end{equation}
where $m_{\rm d}$, $m_{\rm g}$, and $Z$ are the dust mass, gas mass, and metallicity
in a cell.
The density in a dust grain is assumed to be that of silicate-like grains, 3\,g\,cm$^{-3}$.
The optical depth by dust is given by
\begin{equation}
d\tau_{\rm dust} = \int Q(a_{\rm d}, \nu) \pi a^{2}_{\rm d} 
\left( \frac{dn_{\rm d}}{d a_{\rm d}} \right) d a_{\rm d} ds,
\end{equation}
where $Q(a_{\rm d}, \nu)$ and $n_{\rm d}$ are the absorption coefficient factor
and the grain number density, respectively.
We adopt the $Q$-value of silicate grain derived in \citet{DL84}.

\begin{figure}
\begin{center}
\includegraphics[scale=0.45]{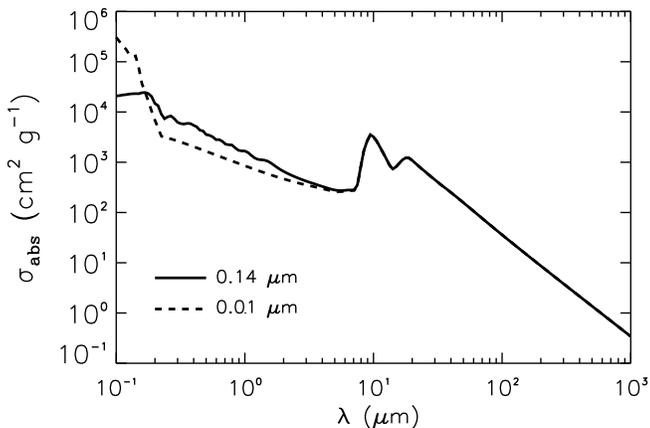}
\caption{
Dust-absorption cross-section per unit dust mass.
Solid and dash lines show the cross section of $0.14~\rm \mu m$ and $0.01~\rm \mu m$ 
silicate dust, respectively. The Q-value of silicate dust is taken from 
\citet{DL84}.
}
\label{fig:mass_sigma}
\end{center}
\end{figure}

Of course, there are some options for the dust model. 
The present dust model is motivated by the supernova dust model in \citet{Nozawa07}.
In their model, the small dust grains of $\lesssim 0.01~\rm \mu m$ are readily destroyed
in shock wave, so that the typical size becomes $\sim 0.1~\rm \mu m$.
On the other hand, \citet{Todini01} pointed out that the typical size 
could be reduced to $\sim 0.01~\rm \mu m$ for the first grains in the early universe. 
The number density of dust grains increases with decreasing dust size
for the constant amount of dust,
while the absorption cross section decreases with the dust size.
In Figure~\ref{fig:mass_sigma}, the extinction per unit dust mass 
is compared between $\sim 0.14~\rm \mu m$ and $\sim 0.01~\rm \mu m$ dust.
As seen in Figure~\ref{fig:mass_sigma}, the absorption efficiency  
does not change appreciably with changing the dust size.
Therefore, the infrared luminosity and dust temperature are not so sensitive to the dust size.
Actually, in a test calculation for $0.01~\rm \mu m$ dust, the temperature decreases by just a few K.
The difference in the peak flux at 850 $\mu$m is a few per cent,
because the peak of thermal emission shifts toward 850 $\mu$m with decreasing temperature.

On the other hand, the Calzetti's law 
\begin{equation}
\frac{A_\lambda}{A_V} = \begin{cases}
0.657 (-1.857 + 1.040 / \lambda) + 1\\
~~~~~~~~~~~~~~~~~~~~{\rm for} ~ 0.63~{\rm \mu m} \leq \lambda \leq 2.20~{\rm \mu m}\\
0.657 (-2.156 + 1.509 / \lambda - 0.198 / \lambda^{2} + 0.011 /  \lambda^{3}) + 1\\
~~~~~~~~~~~~~~~~~~~~{\rm for} ~ 0.12~{\rm \mu m} \leq \lambda < 0.63~{\rm \mu m}, 
\end{cases}
\end{equation}

\begin{figure}
\begin{center}
\includegraphics[scale=0.45]{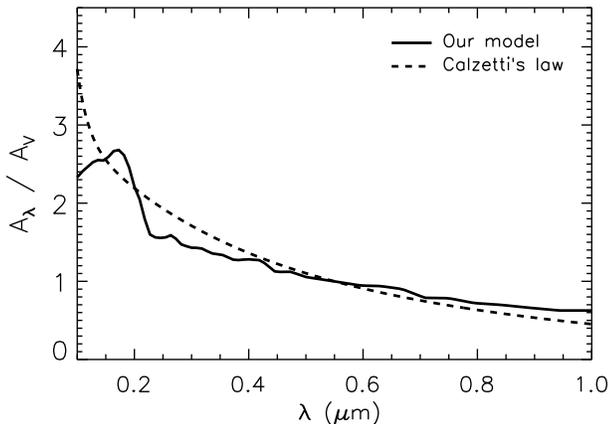}
\caption{
The dust extinction curve.
Solid and dash lines show the dust model in this paper and Calzetti's law \citep{Calzetti00},
respectively. 
}
\label{fig:calzetti}
\end{center}
\end{figure}

\citep{Calzetti00} is frequently used for dust extinction in local galaxies.
In Figure~\ref{fig:calzetti}, the extinction curve of our dust model is compared to the Calzetti's law.
Our model is somewhat flatter than the Calzetti's law in UV-optical range.
However, even if we use Calzetti's extinction curve, the infrared luminosity 
does not significantly change from our model.
Actually, by a test calculation, we find that the relative difference is twenty per cent for IR luminosity, 
and the difference in dust temperature is $\sim$ a few K.

\subsection{Radiative equilibrium}
We evaluate the dust temperature $T_{\rm dust}$ by solving the radiative equilibrium
between heating by photo-absorption ($\Gamma$) and cooling by thermal emission ($\Lambda$),
which are given by
\begin{equation}
\begin{split}
\Gamma &= \int  \int \; 4 \pi J_{\nu} Q(a_{\rm d}, \nu) \pi a_{\rm d}^{2} \frac{dn_{\rm d}}{da_{\rm d}} \; da_{\rm d} \; d\nu + k_{\rm CMB}\\
\Lambda &= \int  \int \; 4 \pi^{2} a_{\rm d}^{2} B_{\nu} (T_{\rm dust}) Q(a_{\rm d}, \nu) \frac{dn_{\rm d}}{da_{\rm d}} \; da_{\rm d} \; d\nu ~,
\end{split}
\end{equation}
where $J_{\nu}$ and $k_{\rm CMB}$ are the mean intensity of stellar radiation 
and the heating term by CMB radiation, respectively.
We simplify the right side of this equation as follows \citep[e.g.,][]{Evans94},
\begin{equation}
\int  \int \; 4 \pi^{2} a_{\rm d}^{2} B_{\nu} (T_{\rm dust}) Q(a_{\rm d}, \nu) \frac{dn_{\rm d}}{da_{\rm d}} \; da \; d\nu
\;\; \sim \;\;
4\pi^{2} \bar{a}_{\rm d}^{2} n_{\rm d} \bar{Q} \sigma T_{\rm dust}^{4},
\end{equation}
where $\sigma$ is the Stefan-Boltzmann constant,
$\bar{a}_{\rm d}$ is the mean dust size weighted by the size distribution function,
\begin{equation}
\bar{a}_{\rm d} = \frac{\int a_{\rm d} \frac{dn_{d}}{da_{\rm d}} da_{\rm d}}
{\int \frac{dn_{\rm d}}{da_{\rm d}} da_{\rm d}} ~.
\end{equation}
The $\bar{Q}$ is a mean $Q$-value weighted by the Planck function,
\begin{equation}
\bar{Q}(T_{\rm dust}) = \frac{\int B_{\rm d}(T_{\rm dust}, \nu) Q(\bar{a}_{\rm d}, \nu) d\nu}
{\int B_{\rm d} (T_{\rm dust}, \nu) d\nu}~.
\end{equation}
Then, we obtain the dust temperature by
\begin{equation}
T_{\rm dust} = \left(
			 \frac{ \Gamma}
			{4\pi^{2} \bar{a}_{\rm d}^{2} \bar{n}_{\rm d} \bar{Q} \sigma}
		  \right)^{1/4}.
\end{equation}

%
%

\section{Results}
\subsection{Evolution of dust component}
\begin{figure*}
\begin{center}
\includegraphics[scale=0.5]{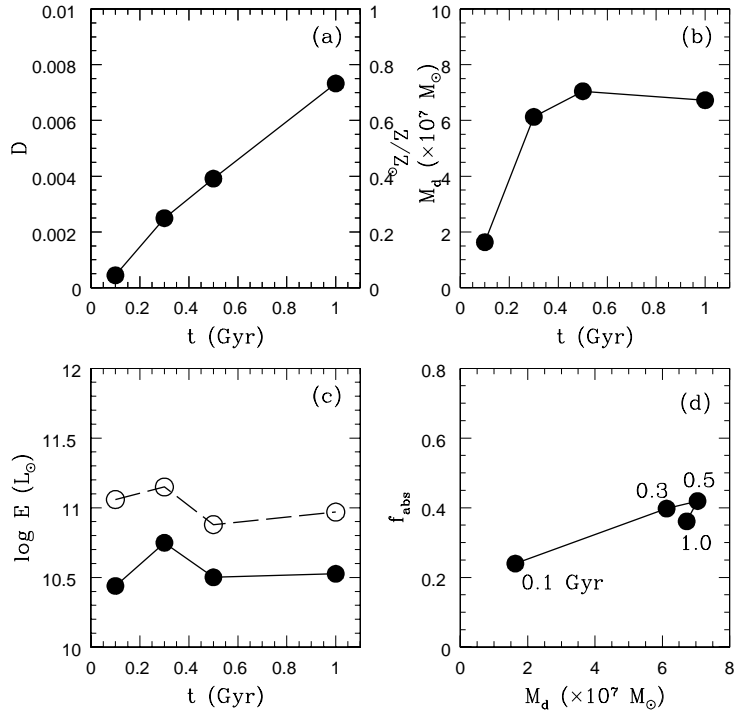}
\caption{
$upper ~left$ : Dust-to-gas mass ratio as a function of evolution time.
$upper~ right$ : Total dust mass as a function of evolution time. The dust mass is estimated 
in proportion to metallicity.
$lower~ left$ : Radiation energy absorbed by dust as a function of evolution time (filled circles).
Open circles are intrinsically radiated energy from stars.
$lower ~ right$ : 
The absorbed fraction to the total radiated energy
as a function of the dust mass.
}
\label{fig: all}
\end{center}
\end{figure*}

In Figure~\ref{fig: all}, we present the time variation of dust component
as a function of the galactic evolution time. 
Figure~\ref{fig: all}(a) shows the dust-to-gas mass ratio ($D$).
The ratio increases monotonically with the galaxy evolution owing to continuous supernovae.
$D$ is $4\times 10^{-4}$ (corresponding to the metallicity $Z = 4 \times 10^{-2} Z_\odot$,
where $Z_\odot$ is the solar metallicity) 
at 0.1 Gyr and reaches $0.7 \times 10^{-2}$ ($Z = 0.7 Z_\odot$) at 1.0 Gyr.

Figure~\ref{fig: all}(b) shows the total dust mass in the model galaxy.  
The dust mass is $1.6 \times 10^{7}~\Msun$ at 0.1 Gyr, and then
$(6-7) \times 10^{7}~\Msun$ at $\geq$ 0.3 Gyr.
In contrast to the monotonic increase of metallicity,
the total dust mass does not change greatly at $\geq$ 0.3 Gyr.
This trend comes from the fact that the continuous energy input 
by multiple supernovae results in the blowout 
of dusty gas from the halo.
The fraction of escaped gas to the initial amount becomes 
$\sim 50$ per cent at $t_{\rm age} = 1.0$ Gyr.

Our model shows that the metallicity reaches the level of 
$0.2 Z_\odot < Z < 0.4 Z_\odot$ in the late phase of LAE.
Very recently, \citet{Nakajima11} assessed 
the metallicity of LAEs to be $Z \gtrsim 0.16 Z_\odot$, which is 
significantly higher than was previously thought for LAEs.
This is concordant with our results.

Figure \ref{fig: all}(c) shows the radiation energy absorbed by dust,
and Figure \ref{fig: all}(d) shows 
the absorbed fraction to the total radiation energy emitted by stars.
The model galaxy has a blue SED owing to young stars, and
the absorption efficiency by dust decreases steeply with increasing wavelength.
Thus, most of energy absorbed by dust is attributed to UV-optical continuum radiation
from stars in the range of $\lambda  \sim 1000 - 6000 ~\rm \AA$, 
and strongly depends on SFR and dust amount.
Figure \ref{fig: all}(c) shows that
the absorbed energy by dust is peaked at 0.3 Gyr,
although the metallicity increases monotonically. 
This is because the dust is distributed compactly around star forming regions 
in an early phase, and hence the UV photons from stars are effectively absorbed.

\subsection{Dust temperature}
\begin{figure*}
\includegraphics[angle=270,scale=0.4]{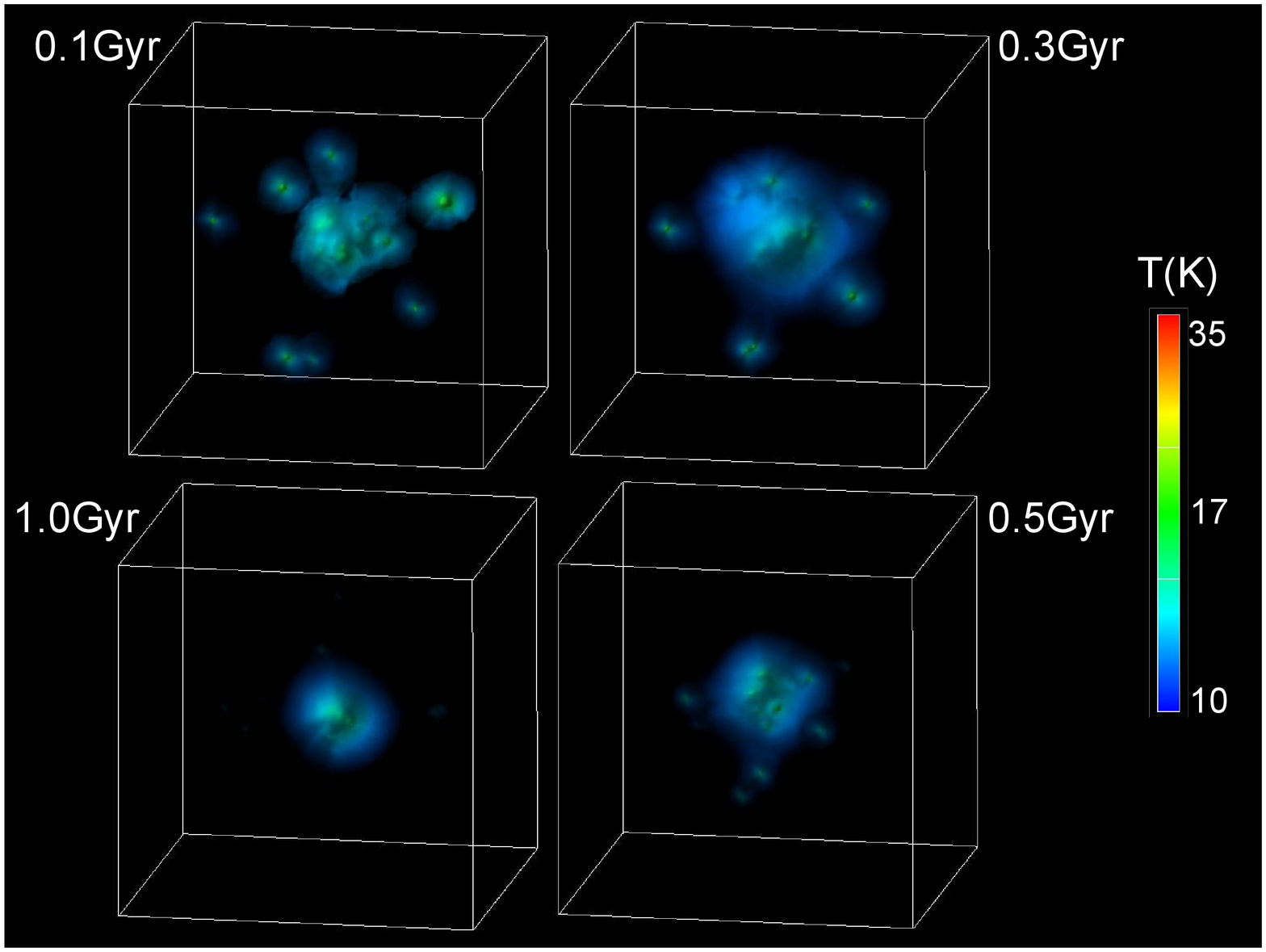}
\caption{
The distributions of dust temperature when the model galaxy is placed at $z=3$.
The color shows dust temperature.
}
\label{fig:tempdist}
\end{figure*}

Figure~\ref{fig:tempdist} shows the dust temperature distributions,
when the model galaxy is placed at $z=3$.
The dust temperature  is $T_{\rm dust} \sim 35~\rm K$ near star-forming regions, while
$T_{\rm dust} \sim 10~\rm K$ far from there.
In an early phase ($t_{\rm age} =$ 0.1 - 0.3 Gyr), the high-temperature regions are distributed 
extensively according to the distributions of star-forming regions.
In a later phase ($t_{\rm age} =$ 0.5 - 1.0 Gyr), the distributions of high-temperature dust 
are confined into the central 
regions as a result of the dynamical evolution of star-forming regions.

\begin{figure}
\begin{center}
\includegraphics[scale=0.45]{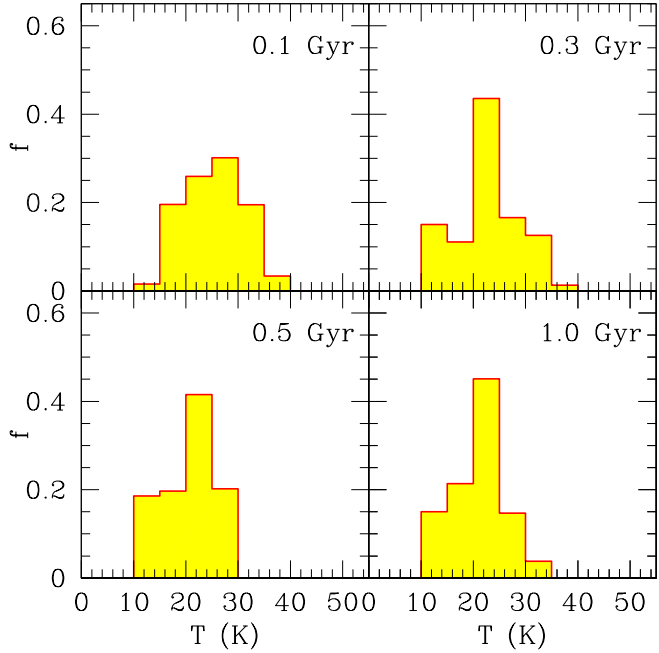}
\caption{
Mass fraction of dust to the total amount in a given temperature range. 
The bin size is $5~\rm K$.
}
\label{fig:tempfrac}
\end{center}
\end{figure}

We statistically study the dust temperature distributions.
Figure~ \ref{fig:tempfrac} shows the mass fraction of dust in a given temperature range.
In an early stage, dust is confined in each sub-halo, and 
therefore it is distributed near star-forming regions.
Thus, at $t_{\rm age} = 0.1$ Gyr, 
a part of dust has high temperature of $T_{\rm dust} \geq 35$ K,
and the mass fraction is broadly distributed in the range of $T_{\rm dust} = 10 - 40$ K.
Thereafter, the dusty gas is blown away by supernova explosions and 
is distributed diffusely in the extended regions. 
Then, a part of dust, which is far from young stars, cannot absorb much radiation energy,
and therefore becomes in a cooler state.
Hence, in a later phase, dust with high temperature of $T_{\rm dust} \geq 35$ K disappears,
and the dust temperature falls in the range of $T_{\rm dust} \approx 10 - 30$ K.

Figure~\ref{fig:tave} shows the mean temperature as a function of evolution time.
Filled and open circles are the mass weighted mean temperatures
and the energy weighted ones, respectively.
The mass weighted mean is $\sim 24~\rm K$ in the LAE phase, while
$\sim 20~\rm K$ in the LBG phase.
The energy weighted mean is $\sim 31~\rm K$ in the LAE phase, while
$\sim 26~\rm K$ in the LBG phase.
The high-temperature dust emits thermal radiation more effectively than colder one 
owing to the higher Q-value and the strong temperature-dependence of emissivity.
Hence, the energy weighted mean temperatures are higher by $\sim \rm 5 K$ 
than the mass weighted ones.
The mean temperatures decline with the evolution time
especially in early phases. This is intimately relevant to the sub-mm brightness
of the galaxy. 
Recently, \citet{Hwang10} 
have studied the dust properties of galaxies with $L_{\rm IR} = 10^{11}~\rm \Lsun$ 
in the redshift range $0.1 \lesssim z \lesssim 2.8$,
and found the dust temperature to be $T_{\rm dust} \sim 20 - 50$ K. 
\citet{Amblard10} have found that the average dust temperature is
$28 \pm 8$ K for submm galaxies with the average redshift of $2.2 \pm 0.6$.
These temperatures are comparable to the dust temperature in our calculations.

\subsection{Spectral Energy Distributions}
We compute the intrinsic spectral energy distributions (SED) 
of the stellar component in the model galaxy
by using the population synthesis code P\'{E}GASE v2.0 \citep{Fioc97}.
We place the model galaxy at several redshifts.
The initial mass function (IMF) is assumed to be that by \citet{Salpeter55}  
in the mass range of $0.1 - 50 \Msun$.
Also, the effects by the age and metallicity of stellar population are incorporated
by interpolating the table generated by P\'{E}GASE.
The dust temerature is primarily determined by the UV continuum from OB stars
with $\gtrsim 2\Msun$.
For the Salpeter IMF in the range of $0.1 - 50 \Msun$, the mass fraction of
stars with $\gtrsim 2\Msun$ is 0.267. If we suppose the range of 
$0.1 - 100 \Msun$, the fraction is 0.287.  
Hence, the upper bound of mass range is not so significant on 
determining the dust temperature.

In this work, at first, we derive the intrinsic SED of stellar component by P\'{E}GASE,
and correct the SED by incorporating the dust extinction through three-dimensional RT simulations.
Then, we evaluate the dust temperature and the thermal emission following the equation (7).
Finally, we get the emergent SED by combining the dust absorption-corrected stellar SED with
the thermal emission from dust.
Fig. \ref{fig:sed} shows the resultant SED of the model galaxy in the observed frame
after making the K-correction.
As well known, $\lambda \gtrsim 1\; \rm mm$, the flux density does not decrease 
with redshift owing to the effect of negative K-correction.

\subsection{Sub-millimeter brightness}
Here, we see the time variations of the sub-millimeter brightness,
focusing on the flux at $850~\rm \mu m$.
Figure~\ref{fig:smtotal} shows the total flux at $850~\rm \mu m$ 
in observed frame ($S_{850}$).
Interestingly, the sub-millimeter brightness is peaked at 0.3 Gyr (LAE phase), 
independent of the assumed redshift. 
As already shown in Figure~\ref{fig: all}(a), the dust-to-gas ratio
(metallicity) increases monotonically with time. But, the sub-millimeter brightness 
does not increases according to the increase of metallicity.
This is because, in the LAE phase, dust grains are concentrated on clumpy 
star-forming regions and therefore the dust grains can be effectively 
heated up by stellar radiation. On the other hand, 
the continuous supernovae drive dust grains far away 
from star-forming regions at $\geq 0.5$ Gyr (LBG phase). 
Then the grains cannot absorb much radiation from stars
and turn into a cold state. As a result, the sub-millimeter brightness declines.

\begin{figure}
\begin{center}
\includegraphics[scale=0.4]{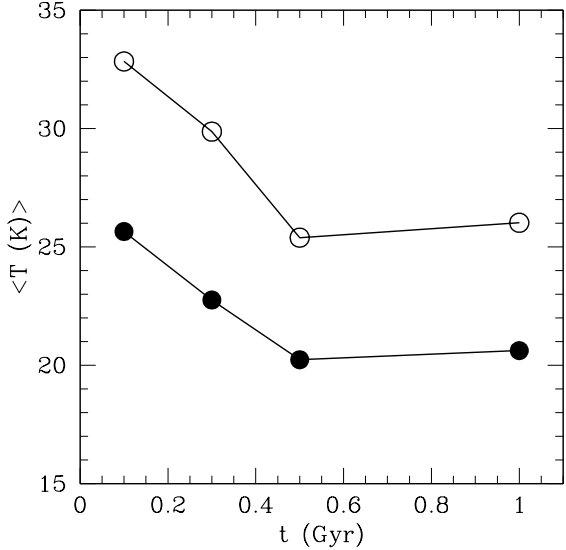}
\caption{
The mean temperature of dust as a function of evolution time.
Open and filled circles are the energy weighted mean
and the mass weighted mean, respectively.
}
\label{fig:tave}
\end{center}
\end{figure}

\begin{figure}
\begin{center}
\includegraphics[scale=0.45]{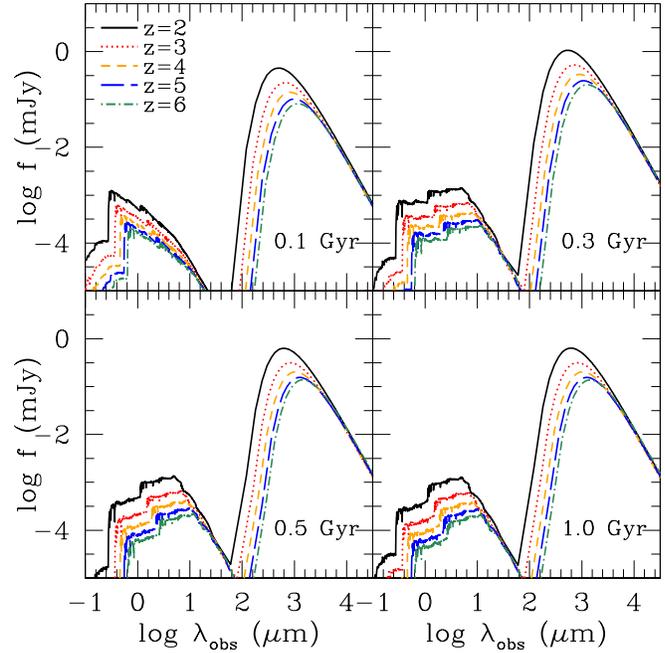}
\caption{
Spectral energy distributions of the model galaxy in the observed frame
after the K-correction is made. 
Different line style corresponds to a different redshift at which the model galaxy is placed
(solid: $z=2$, dot: $z=3$, dash: $z=4$, long-dash: $z=5$ and dot-dash: $z=6$).
}
\label{fig:sed}
\end{center}
\end{figure}

\begin{figure}
\begin{center}
\includegraphics[scale=0.4]{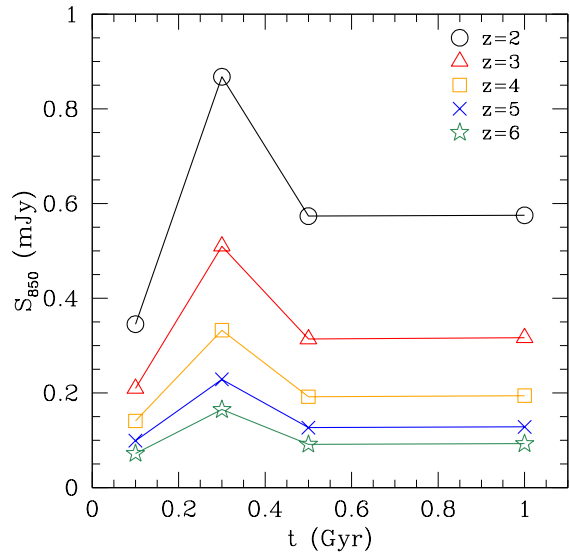}
\caption{
Total flux at $850 ~\rm \mu m$ in the observed frame as a function of evolution time.
We set the model galaxy at each redshift 
(circle: $z=2$, triangle: $z=3$, square: $z=4$, cross: $z=5$ and star: $z=6$).
}
\label{fig:smtotal}
\end{center}
\end{figure}

The relative sub-millimeter flux at each epoch depends on the redshift $z$.
If the galaxy is at $z \lesssim 3$, the total flux at 0.1 Gyr is smaller than that at 0.5 Gyr and 1.0 Gyr,
although the total absorbed energy is nearly equal.
This is due to the fact that the flux of $850~\rm \mu m$ is sensitive to the dust temperature.
The spectral shape of dust emission is basically that of the black body, 
although the $Q$-value affects the shape to some degree.
The temperature, which exhibits a peak at $850~\rm \mu m$ in the observed frame, is
$13.65~\rm K$ for a galaxy at $z=3$.
As seen in Figure~\ref{fig:tave},
the mass-weighted mean temperature declines with the evolution time,
and the temperature at 0.5 Gyr and 1.0 Gyr is closer to 13.65 K than that at 0.1 Gyr.
Therefore, we conclude that the low sub-mm flux in an earlier LAE phase
is not attributed to the low dust amount but rather to the higher dust temperature.
For higher galaxy redshifts, 
the absolute flux decreases of course with increasing redshift.
But, the difference of the flux between $0.1$ and 0.5 (or 1.0) Gyr becomes smaller,
since the $850~\rm \mu m$ in the observed frame 
is corresponding to shorter wavelengths.

\begin{figure*}
\begin{center}
\includegraphics[scale=0.6]{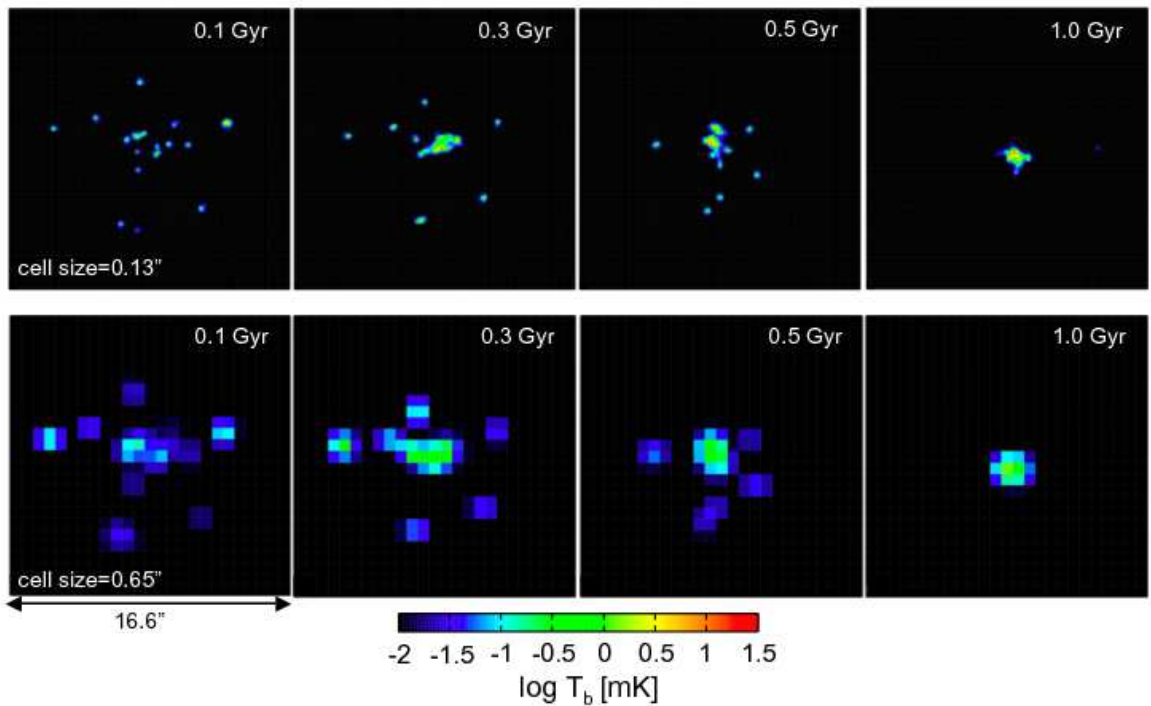}
\caption{
The map of brightness temperature at $\lambda=850~\mu {\rm m}$ in the model galaxy, 
assuming the galaxy to be located at $z=3$. The linear scale of each panel is $16.6 "$.
Upper panels show the map with the original spatial resolution which corresponds
to the angular resolution of $0.13"$.
Lower panels are the map coarse-grained with the angular 
resolution of $0.65"$.
}
\label{fig:smdist}
\end{center}
\end{figure*}

%
%
\section{Detectability by ALMA}

Here, we assess the feasibility to detect early star-forming galaxies 
with SFR of $\sim 10\;\rm \Msunyr$
by {\it Atacama Large Millimeter/submillimeter Array} (ALMA). 
In the present simulation, the total flux of $850 ~\rm \mu m$ in the LBG phase is 
$S_{850}=0.65\sim 0.3$ mJy
for a galaxy at $z=3$. 
Some lensed LBGs have been detected at sub-millimeter,
and show $S_{850}=0.40$ mJy and $SFR \sim 10~\Msunyr$ 
at $z=2.9$ \citep[MS0451-a : ][]{Borys04},
$S_{850}=0.39$ mJy and $SFR \sim 24~\Msunyr$ \citep[cB58 : ][]{Baker04a},
or $S_{850}=0.65$ mJy and $SFR \sim 9~\Msunyr$ at $z=2.5$ \citep[SMMJ16359 : ][]{Kneib05}.
Our results match these observations well.
Also, \citet{Stanway10} showed the upper limit of sub-millimeter flux, $S_{\rm 870} < 0.85~ \rm mJy$,
for LBGs at $z=5$. Our prediction is under their upper limit, and therefore 
higher sensitivity is required to detect sub-mm flux from LBGs at $z \gtrsim 5$.
On the other hand, there are a few bright LBGs detected at sub-millimeter
which show $S_{850} \sim 1-5$ mJy and $SFR \sim 100 - 200~\Msunyr$ \citep{Belokurov07, Coppin07, Chapman09}.
These may be more massive and dusty star-burst systems.

In Fig. \ref{fig:smdist}, the map of the brightness temperature
at $\lambda=850~\mu {\rm m}$ is shown for the model galaxy placed at $z = 3$.
The brightness temperature $T_{\rm b}$ is given by 
\begin{equation}
T_{\rm b} = 11.2
  \left( {\lambda \over 850\mu {\rm m}} \right)^{2}
  \left( {\theta^2 \over 1"\times1"} \right)^{-1}
  \left( {F \over {\rm mJy}} \right) {\rm mK}.
\end{equation}
The higher $T_{\rm b}$ traces the star-forming regions.
Clumpy clouds with higher $T_{\rm b}$ are distributed over an extended area
in the LAE phase ($t_{\rm age} = 0.1 - 0.3$ Gyr),
while those are concentrated near the center 
in the LBG phase ($t_{\rm age} = 0.5 - 1.0$ Gyr).

The upper panels in Figure~\ref{fig:smdist} show $T_{\rm b}$ 
in the original resolution of simulations that corresponds to $\theta \sim 0.13"$. 
The model galaxy exhibits $T_{\rm b} \sim 0.1 - 40$ mK, where
the high $T_{\rm b}$ regions are of $\gtrsim 20$ mK.
The integration time required to detect extended sources of $\sim20$ mK 
with the angular resolution of $\theta \sim 0.13"$ is assessed to be $\sim 5$ hours per each beam.
The sub-millimeter flux strongly correlates with the stellar distribution.
At an early phase, the area to cover the half mass of stars is $\sim 13~\rm arcsec^{2}$.
At a later phase, the model galaxy becomes as compact as $\sim 4 ~\rm arcsec^{2}$.
Hence, the number of beams to cover the area will be quite large.
Even for the later compact phase, the required time to cover the area will be a few hundred hours.
Thus, the observations with the angular resolution of $\sim 0.13"$ do not seem achievable 
to trace the sub-mm structure for early star-forming galaxies.

The lower panels of Figure~\ref{fig:smdist} show $T_{\rm b}$ in the coarse-grained 
resolution of $\theta \sim 0.65"$.
The high $T_{\rm b}$ regions show  $\gtrsim 3$ mK.
In such an angular resolution, the sensitivity with 16 antennas of ALMA
is $\sim 2$ mK by $\sim 60$ minute integration.
If an early star-forming galaxy is observed, some beams may detect sub-mm flux 
by the integration of less than 1 hour.
Thus, the detection of clumpy sub-mm features of an early star-forming galaxy seems feasible.
Moreover, observations with lower angular resolution of  $\gtrsim 1"$ allow us
to study the statistics of high-$z$ sub-mm sources, e.g., the luminosity function of sub-mm galaxies.
ALMA will detect the sub-millimeter flux from LBGs at $z \sim 3$ by $\sim 10$ minute integration with 16 antennas.
Even for a galaxy at $z = 6$, ALMA can detect it by $\sim 60$ minute integration.

Recently, Dayal et al. (2011) have estimated a bit lower flux at
$850~\rm \mu m$ than the present prediction. 
They have assumed homogeneous distributions of stars and dust 
for an LAE identified in a cosmological simulation.
Compared to our results, it implies that the clumpiness of star-forming regions
and interstellar medium significantly contributes
to the enhancement of the sub-mm brightness.


\section{SUMMARY}

We have performed three-dimensional radiative transfer calculations on
high-resolution hydrodynamic simulations with inhomogeneous metal enrichment.
Then, we have explored the dust temperature and sub-mm flux in a high-redshift 
star-forming galaxy.
Attention has been concentrated on the sub-mm properties 
of LAE and LBG phases. 
The star formation rate (SFR) is $\sim 10 \rm\; \Msunyr$ in the 
LAE phase, and several $\rm\; \Msunyr$ in the LBG phase.
As a result, we have found that dust grains concentrated on clumpy 
star-forming regions can effectively absorb UV radiation from stars
in the LAE phase, and then the grains are heated up to $T_{\rm dust} \gtrsim 35$ K.
On the other hand, in the LBG phase, the continuous supernovae blow away dusty gas 
from star-forming regions. Hence, the grains cannot absorb much radiation from stars
and turn into a cold state.
Resultantly, the emergent sub-millimeter brightness is peaked in an LAE phase
around $t_{\rm age}\sim0.3$ Gyr, independent of the assumed redshift. 
This shows that the sub-millimeter brightness does not necessarily increases 
according to the increase of the dust-to-gas ratio or the metallicity. 
We have found that the clumpiness of star-forming regions
and interstellar medium significantly enhances the sub-mm brightness.

Also, by deriving the spectral energy distributions, we have assessed 
the flux at 850 $\rm \mu m$. The flux is found to be $S_{850} \sim 0.2 - 0.9$ mJy, 
if the model galaxy is placed at $6 \geq z \geq 2$.
Even for $z = 6$, the sub-mm flux does not decrease largely because of negative K-correction.
With angular resolution of $\theta \sim 0.65"$, ALMA can detect
the sub-mm flux from an LAE as well as an LBG by the integration of less than 1 hour
with 16 antennas.
Observations with angular resolution of  $\gtrsim 1"$ allow us to detect
high-z sub-mm sources by $\sim 10$ minute integration with 16 antennas of ALMA. 
Therefore, we can study clumpy sub-mm features of an early star-forming galaxy and
also the statistics of the luminosity function of sub-mm galaxies with ALMA.

%
%
\section*{Acknowledgments}
We are grateful to K. Nagamine, A. Inoue, Y. Miyamoto and M. Ouchi
for valuable discussion and comments. 
We thank the anonymous referee for useful comments.
Numerical simulations have been performed with the {\it FIRST} simulator  
and {\it T2K-Tsukuba} at Center for Computational Sciences, in University of Tsukuba. 
This work was supported in part by the {\it FIRST} project based on
Grants-in-Aid for Specially Promoted Research by 
MEXT (16002003) and
JSPS Grant-in-Aid for Scientific Research (S) (20224002),
(A) (21244013), and (C) (18540242).

%
%




\label{lastpage}

\end{document}